\documentclass[conference]{IEEEtran}
\usepackage{cite}
\usepackage[cmex10]{amsmath}
\usepackage{amssymb}
\usepackage{url}
\usepackage{subfig}
\usepackage{algorithmic,algorithm}
\usepackage{bigdelim,multirow}
\usepackage{graphicx}

\begin{document}
%
\title{Reflective Network Tomography\\Based on Compressed Sensing}

\author{\IEEEauthorblockN{Kensuke Nakanishi\IEEEauthorrefmark{1},
Shinsuke Hara\IEEEauthorrefmark{1}\IEEEauthorrefmark{3},
Takahiro Matsuda\IEEEauthorrefmark{2}\IEEEauthorrefmark{3},
Kenichi Takizawa\IEEEauthorrefmark{3},
Fumie Ono\IEEEauthorrefmark{3}, and
Ryu Miura\IEEEauthorrefmark{3}
}

\IEEEauthorblockA{\IEEEauthorrefmark{1}Graduate School of Engineering,
Osaka City University\\Osaka, 5588585, Japan\\
Email:\{nakanishi.k@c., hara@\}info.eng.osaka-cu.ac.jp}
\IEEEauthorblockA{\IEEEauthorrefmark{2}Graduate School of Engineering, Osaka
University\\Osaka, 5650871, Japan\\
Email: matsuda@comm.eng.osaka-u.ac.jp}
\IEEEauthorblockA{\IEEEauthorrefmark{3}Wireless Network
Research Institute, National Institute of Information and Communications
Technology (NICT)\\
Kanagawa, 2390847, Japan\\
Email:\{takizawa, fumie, ryu\}@nict.go.jp}
}

\maketitle

\begin{abstract}
Network tomography means to estimate internal
link states from end-to-end path measurements.
In conventional network tomography,
to make packets {\em transmissively} penetrate a network,
a cooperation
between transmitter and receiver nodes
is required, which are located at different places in the network.
In this paper, we propose a {\em reflective network tomography},
which can
totally avoid such a cooperation, since
a single transceiver node transmits packets and receives them after traversing back from the network.
Furthermore, we are interested in identification of a limited number of bottleneck links, so
we naturally introduce compressed sensing technique into it.
Allowing two kinds of paths such as (fully) loopy path and folded path,
we propose a computationally-efficient algorithm for constructing
reflective paths for a given network.
In the performance evaluation by computer simulation,
we confirm the effectiveness of the proposed reflective network tomography scheme.

\end{abstract}


\IEEEpeerreviewmaketitle

\section{Introduction}
Tomography refers to the cross-sectional imaging of an object from either
{\em transmission} or {\em reflection} data collected by illuminating
the object from
many different directions~\cite{Kak}.
When the object is an information network,
it is called {\em network tomography}~\cite{Vardi1996},
which has been used
to encompass a class of approaches to infer the internal
link states from end-to-end path measurements~\cite{Coates2002}.
The end-to-end path behaviors have been {\em transmissively} measured via a cooperation between transmitter and receiver nodes, which are located at different places in a network.
However if it is possible to eliminate such a cooperation,
network tomography would become a more powerful method with special properties
({\em implementability}, {\em adaptability} and {\em asynchronism})
for measuring and analyzing network specific characteristics.

In this paper,
according to the types of end-to-end path measurements acquisition,
we first classify network tomography into {\em transmissive} and {\em
reflective} network tomography,
and after discussing their characteristics,
we propose a new reflective network tomography scheme.
Here, in the reflective network tomography scheme,
we focus only on identification of a limited number of links with large delays in a network,
where such links are referred to as {\em bottleneck links}.
In this scheme,
a node acts as both a transmitter and a receiver, i.e., as a transceiver:
it transmits multiple packets over a network along pre-determined
different paths and receives the packets after they traverse back
from the network.
On the other hand,
network tomography is formulated as
an undetermined linear inverse problem and it cannot be always solved.
However, the assumption in the bottleneck link identification
makes it possible to use compressed sensing technique.
To propose the new reflective network tomography scheme,
we tackle two problems:
how to formulate the tomography scheme
and how to determine
{\em going around} paths from/to a transceiver node.


Note that,
although end-to-end path measurements can be conducted either actively or passively,
reflective network tomography scheme is only based on active measurements.
Thus, we particularly consider active tomographic scheme in this paper.

\section{Network Tomography}
\label{sec:tomography}

\subsection{Transmissive Network Tomography}
\begin{figure}[ttt]
\centering
\includegraphics[width = .68\linewidth ]{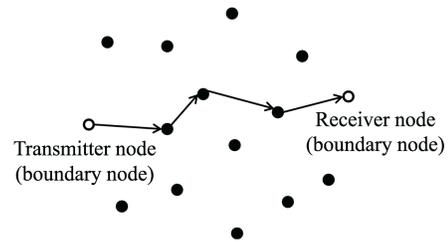}
\caption{Transmissive end-to-end path measurement.}
\label{TEM}
\end{figure}

In this subsection we define transmissive network tomography
via some
examples~\cite{Cascares1999,Coates2000,Bu2002,Takemoto2013,Firooz2014,Duffield2000}
which are characterized by {\em transmissive end-to-end path measurements}.
Fig.~\ref{TEM} shows an example of a transmissive end-to-end path measurement~\cite{Takemoto2013}.

In a network with a defined boundary,
it is assumed that access is available to nodes
at the boundary, but not to any in the interior.
In order to get transmissive end-to-end path measurements,
some boundary nodes are selected as {\em transmitter} and {\em receiver}
nodes.
For example, in \cite{Takemoto2013},
two nodes are respectively assigned
as a transmitter and a receiver,
whereas in \cite{Firooz2014}, there are many transmitter and receiver nodes.
The transmitter nodes
send probe packets to all (or a subset of) the receiver nodes
to measure packet attributes on the paths between them.
Accordingly, each probe packet transmissively penetrates the network
along a {\em measurement path},
and brings a transmissive end-to-end path measurement.
In \cite{Coates2000}, a transmissive tomographic methodology based on
unicast communication is proposed.
In \cite{Cascares1999} and \cite{Bu2002},
on the other hand,
a single-source multicast transmission
by a single or multiple transmitter nodes is applied to networks
with tree and general topologies, respectively.
From such transmissive end-to-end path measurements between transmitter and
receiver nodes,
the internal network states such as link-level network parameters can be
estimated.
For example, in \cite{Duffield2000},
link delay variance is estimated from transmissive end-to-end path
measurements in a multicast setting.

\subsection{Reflective Network Tomography}
Unlike transmissive network tomography,
reflective network tomography eliminates the need for special-purpose
cooperation from receiver nodes.
Namely, an end-to-end path measurement is calculated from records on only one
node.
A boundary node is selected as a {\em transceiver} node,
and it injects probe packets into the network.
Each probe packet goes and back to the transceiver node along a
different measurement path,
and brings {\em reflective end-to-end path measurements}.
For example, in \cite{Tsang2004},
a reflective network tomography scheme based on round trip time (RTT)
measurement
only along a folded path (see \ref{sec:system} for its definition)
is proposed to estimate the delay variance for a link of interest.
Thus,
in contrast to transmissive network tomography,
reflective network tomography is defined by
{\em reflective end-to-end path measurements}.

\section{Properties of Reflective Network Tomography}
\label{sec:properties}
\subsection{Implementability}
The methods described in the above transmissive network tomography all
require a coordination between transmitter and receiver nodes.
However, the following problems have not been discussed deeply:
how to access
all the transmitter and receiver nodes
and how to establish the coordination between them,
in order to implement the network tomography, i.e.,
designate the measurement paths, transmit active probe packets
and collect the end-to-end path measurements.
In a network,
these would occupy some part of the time/frequency resource
and consume some energy.
Without any solution strategy,
these problems would limit the scope of the paths over which the
measurements can be made.
Thus most of them would not be widely applicable because of the lack of
an available widespread
infrastructure for transmissive end-to-end path measurements.

On the other hand,
the reflective network tomography scheme does not require special
cooperation from the other interior and boundary nodes,
because the reflective end-to-end path measurements are calculated only by a
single transceiver node.
We just use the transceiver node to implement the reflective network
tomography,
so we can say that the reflective network tomography can be carried out
more easily.

\subsection{Adaptability}
Most of the existing transmissive network tomography schemes are
based on non-adaptive measurements in themselves.
Namely, the measurement paths are often fixed in advance and
do not depend on the previously acquired measurements.
The reason is that it is difficult to feed back the prior end-to-end path measurements
from receiver nodes to transmitter nodes every probing.

In the reflective network tomography scheme, on the other hand,
since the probe packets return to the transceiver node,
measurement paths can be adaptively selected
depending on the previously gathered information.
So it can give us the advantage of sequential
measuring schemes that adapt to network states using information
gathered throughout a measurement period.
Furthermore, many current methodologies usually assume that
network states are stationary throughout the tomography period.
Even when this assumption is not satisfied, however,
reflective network tomography scheme may be workable
thanks to its adaptability.

\subsection{Asynchronism}
When focusing on transmissive delay tomography which is
transmissive network tomography for link delays,
end-to-end path measurements are usually calculated from the transmission
time and reception time
reported by the transmitter and receiver nodes, respectively.
Therefore, it requires clock synchronization between them.
However, the clock
synchronization is sometimes hard to achieve or not guaranteed,
especially in wireless networks such as wireless sensor
networks, in which electronic components of nodes are too
untrustable to meet the requirement of clock synchronization
in terms of accuracy and complexity~\cite{Bharath2005,Akyildiz2005}.
So, although delay tomography scheme workable in clock-asynchronous networks
is preferable,
to the best of the authors' knowledge,
the transmissive synchronization-free network tomography has been
studied only in \cite{Nakanishi2014}.

On the other hand,
reflective network tomography scheme does not require any clock
synchronization
for any other nodes in a network.
The time delay for a packet traveling through a measurement path
can be estimated
by checking the transmission time and reception time on a transceiver
node's clock.
Therefore, the reflective network tomography scheme is potentially available
in clock-asynchronous networks.

\section{Proposed Reflective Network Tomography Scheme}
\label{sec:formulation}
\subsection{Compressed Sensing}
Compressed sensing is an effective theory in signal/image processing for
reconstructing a finite-dimensional sparse vector based on its
linear measurements of dimension smaller than the size of the
unknown sparse vector~\cite{Donoho2006,Eldar2012}.
Recently, compressed sensing has been also used for network
tomography~\cite{Xu2011,Takemoto2013,Firooz2014}.
In this subsection, as the preliminary for compressed sensing,
we give several definitions.

First, we define the $\ell_p$ norm ($p \geq 1$) of
a vector $\mathbf{x} = [x_1~x_2~\cdots~x_J]^\top \in \mathcal{R}^{J}$
as
\begin{equation}
\| \mathbf{x} \|_p = \Bigl( \sum_{i=1}^{J} |x_i|^p \Bigr)^{\frac{1}{p} },
\end{equation}
where $\top$ denotes the transpose operator.

Next,
we assume that,
through a matrix $\mathbf{A} \in \mathcal{R}^{I \times J}$ ($I <J$),
we obtain a linear measurement vector
$\mathbf{y} = [y_1~y_2~\cdots~y_I]^\top \in \mathcal{R}^{I}$
for a vector $\mathbf{x} = [x_1~x_2~\cdots~x_J]^\top \in \mathcal{R}^{J}$
as $\mathbf{y}=\mathbf{A}\mathbf{x}$.
Whether or not one can recover a sparse vector $\mathbf{x}$ from
$\mathbf{y}$ by means of compressed sensing
can be evaluated by
the mutual coherence $\mu(\mathbf{A})$~\cite{Eldar2012}.
To calculate the mutual coherence of $\mathbf{A}$,
by picking up the {\it j}-th and {\it j'}-th column vectors from
$\mathbf{A}$
we construct the partial matrix as
\begin{equation}
\mathbf{A} _{jj'} =[\mathbf{c}_j~\mathbf{c}_{j'} ]
\label{original},
\end{equation}
where
$\mathbf{c}_j$ and $\mathbf{c}_{j'}$ are
the $j$-th and $j'$-th column vectors of $\mathbf{A}$, respectively.
The mutual coherence $\mu(\mathbf{A})$
is defined as the maximum value of
$\nu(\mathbf{A}_{jj'})$
($1 \leq j,j' \leq J,j \neq j'$):
\begin{eqnarray}
\mu(\mathbf{A}) & = & \max_{1 \leq j,j' \leq J,j \neq j'}
\nu (\mathbf{A}_{jj'}),
\label{eqn:mutualcoherence} \\
\nu (\mathbf{A}_{jj'}) & = &
\frac
{ |\mathbf{c}_j^\top \mathbf{c}_{j'} | }
{ \| \mathbf{c}_j \|_2 \| \mathbf{c}_{j'} \|_2}.
\label{eq:normalized}
\end{eqnarray}
If
\begin{equation}
k < \frac{1}{2} \Bigl( 1 + \frac{1}{\mu(\mathbf{A})} \Bigr),
\label{k}
\end{equation}
then there exists at most one vector $\mathbf{x}$ with at most {\it k}
nonzero components that $\mathbf{y} = \mathbf{A}\mathbf{x}$.

\begin{table*}[!t]
\begin{center}
\caption{Symbols in Algorithm~\ref{algorithm}}
\begin{tabular}{|c|l|}\hline
$\mathcal{P}_{{\rm disjoint}}$&Set of node-disjoint paths.\\ \hline
$\overline{ \mathcal{P} }_{{\rm disjoint}}$&Set of
node-disjoint reverse paths, which are constructed
by reversing directions of paths in $\mathcal{P}_{{\rm disjoint}}$\\ \hline
$\mathcal{W}$&Set of definitive measurement
paths.\\\hline
$\mathit{path}^{(a)} + \mathit{path}^{(b)}$&Path
connecting $\mathit{path}^{(a)}$ and $\mathit{path}^{(b)}$.\\ \hline
$\mathcal{P}_{{\rm all}}$&Set of all candidates for
measurement paths.\\ \hline
$F_{\mu} (\mathbf{A})$&Function which returns the
mutual coherence of $\mathbf{A}$ if no column vector equals
$\mathbf{0}$, and number greater than $1$ otherwise.\\ \hline
getCostMin($\mathcal{P}$)&Function which returns a
path whose cost is the minimum
in a path set $\mathcal{P}$. \\ \hline
\end{tabular}
\label{parameters}
\end{center}
\end{table*}
\subsection{System Model}
\label{sec:system}
We consider a delay tomographic scheme which identifies a few bottleneck links
in an asynchronous network from reflective end-to-end path measurements.
Our approach employs unicast communication.
Let $\mathcal{G} = (\mathcal{V}, \mathcal{E})$ denote an
undirected network\footnote{Actually our proposed scheme can also be
extended to directed graph models.}, where $\mathcal{V}$ is the node set,
and $\mathcal{E} \subseteq \mathcal{V} \times \mathcal{V}$ is the link set.
Note that $(i,j) \in \mathcal{E}$ implies that $(j,i) \in \mathcal{E}$
since the graph is undirected.
We assume that the topology is fixed throughout the measurement period
and there is only one transceiver node $s$.

\begin{figure}[ttt]
\centering
\includegraphics[width = .50\linewidth ]{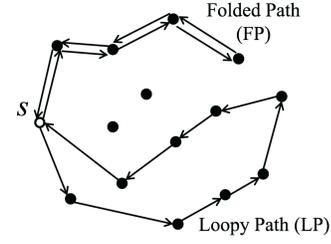}
\caption{Measuring PTTs based on $s$.}
\label{PTTs}
\end{figure}

Due to the fact that the overall delay of a path is the sum of the
delays of all links belonging to the path, delay tomography
problem can be formulated as an inverse problem to recover link
delays based on linear measurements.
Here, we measure {\em packet traveling times} (PTTs)
along two kinds of paths
by injecting probe packets into the network.
One is a (fully) loopy path (LP) defined as the one where any nodes do
not appear more than once
except for a transceiver node,
and the other is a folded path (FP) defined as the one
where any nodes appear twice except for a destination node
(in other words, an FP is composed of a path from a transceiver node to
a destination node
and a path from the destination node to the transceiver node along the
same undirected curve
between them).
Fig.~\ref{PTTs} shows an example for LP and FP,
and we do not consider any path containing partial loops
({\em routing constraint}).
Now, we define $\mathcal{W} = \{\mathit{path}_{s}^{(l)} |~l = 1, 2, \ldots,
|\mathcal{W}|\}$ as a subset of all paths to measure PTTs
based on $s$,
where $\mathit{path}_{s}^{(l)} = \{(s, v^{(l,1)}),
(v^{(l,1)}, v^{(l,2)}), \ldots,
(v^{(l,|\mathit{path}_{s}^{(l)}|-1)}, s)\} \subset
\mathcal{E}$
represents the $l$-th path in $\mathcal{W}$ and
$v^{(l,m)}\in \mathcal{V} \setminus \{s\}$
$(m=1,\ldots,|\mathit{path}_{s}^{(l)}|-1)$
are intermediate nodes in the path.

We reformulate $\mathcal{W}$ and $\mathcal{E}$ as $\mathcal{W} =
\{w_1, w_2, \ldots, w_I\}$ and $\mathcal{E} = \{e_1,
e_2, \ldots, e_J\}$, respectively, where $I = |\mathcal{W}|$ and $J =
|\mathcal{E}|$ denote the numbers of paths and links,
respectively.
We assume that link delays $d_{e_j}$ arise independently on each link
$e_j$ ($j = 1,2, \ldots, J$),
which does not depend on the direction.
Thus, a probe packet transmitted on a path $w_i$ ($i = 1,
2, \ldots, I$) is successfully returned to $s$ with total delay
$D_{w_i}= \sum_{e_j \in w_i} d_{e_j}$.
We define {\em measurement vector} $\mathbf{y} =
[y_1~y_2~\cdots~y_I]^\top$ and {\em link delay vector} $\mathbf{x} =
[x_1~x_2~\cdots~x_J]^\top$~as
\begin{equation}
\begin{split}
y_i &=D_{w_i}= \sum_{e_j \in w_i} d_{e_j},\\
x_j &=d_{e_j}.
\end{split}
\end{equation}
Then, we obtain
\begin{equation}
\label{eqn:system}
\mathbf{y} = \mathbf{Ax},
\end{equation}
where $\mathbf{A} \in \{0,1,2\}^{I \times J}$ represents the ({\em
reflective}) {\em routing matrix} of
$\mathcal{W}$, i.e.,
$(i, j)$-th component
$a_{ij}$ ($i = 1, 2, \ldots,
I,~j = 1, 2, \ldots, J$) in $\mathbf{A}$ is set to $a_{ij} = 1$ or
$a_{ij} = 2$ if $e_j \in w_i$,
and $a_{ij} = 0$ otherwise.
The row size $I$ is related to the interval devoted to the tomography scheme,
and the {\em entrywise matrix norm} of $\mathbf{A}$
\begin{equation}
\| \mathbf{A} \| = \Bigl( \sum_{i=1}^{I} \sum_{j=1}^{J} |a_{ij}| \Bigr)
\end{equation}
is related to the traffic load of probe packets.
The $I$ and $\| \mathbf{A} \|$ determine
the energy required for accomplishing a tomography scheme,
so the former is referred to as {\em the interval factor},
whereas the latter {\em the traffic factor}.
For a given detectability of bottleneck links,
the two factors of a better routing matrix should be smaller.

Note that link states are assumed to be stationary, i.e., link
delays do not change while the proposed scheme is applied,
and a few bottleneck links exist in the network.
Next,
because it is possible to approximate
the elements of $\mathbf{x}$ corresponding to small link delays to be
zero by attributing the delays only to the few bottleneck links,
the idea of compressed sensing can be naturally introduced to network
tomography.
So we utilize compressed sensing based on $\ell_1$-$\ell_2$
optimization~\cite{Zibulevski2010,Matsuda2011}
in order to reduce traffic load of probe packets.
Finally,
when using the PTTs, the assumption of the undirected graph
may lead to inaccurate estimates given the asymmetric
communication~\cite{Zhao2003}.
However, we are interested in identification of a limited number
of bottleneck links,
thus, the assumption can be considered to be valid,
since it does not require measurements with accuracy.

\subsection{Routing Matrix Construction}
\label{sec:algorithm}
Now, we propose a simple algorithm composed of two steps for
constructing a routing matrix $\mathbf{A}$.
This algorithm is for a reflective routing matrix
which can quickly identify a bottleneck, assuming that a bottleneck
link rarely arise in the network.
Algorithm~\ref{algorithm} shows the algorithm,
and Table~\ref{parameters} describes symbols used in
Algorithm~\ref{algorithm}.


First, in STEP~$1$ the algorithm constructs a set of paths as
measurement path candidates
based on {\em node-disjoint paths algorithm} described
in~\cite{Bhandari1999}.
The function NodeDisjointAlgorithm($s$,$v$) in Algorithm~\ref{algorithm}
returns the maximum set of node-disjoint paths from $s$ to $v$.
The set of node-disjoint paths implies the shortest combination of paths
where no nodes are shared among the paths.
By connecting every node-disjoint path from $s$ to $v$ (for all $v \in
\mathcal{V} \setminus s$),
this algorithm lists up the candidates for measurement paths,
which satisfy the routing constraint
that any path is an LP or an FP.

Then, out of the path candidates constructed by STEP 1, STEP~$2$ selects
paths
as measurement paths one-by-one according to the cost of
candidates
until the mutual coherence of the constructed routing matrix becomes
less than 1.0.
If several paths have the same minimum cost, the shortest path is
selected out of them.
Here,
we define the cost function for a measurement path ($\mathit{path} \in
\mathcal{P}_{{\rm all}}$) as
\begin{eqnarray}
\lefteqn{ \mbox{Cost}(\mathit{path}) }
\nonumber \\[6pt]
& = &
\nonumber
\begin{cases}
~\left( \mbox{Number of unused links in $\mathcal{W}$ out of
$\mathit{path}$} \right)^{-1}\\
~~~~~~~~~~~~~~~~~~~~~~~~~~~~~~~~~\left( \mbox{if }F_{\mu}(\mathbf{A}) >
1.0 \right)\\[6pt]
\mbox{Number of } \nu(\mathbf{A}'_{jj'}) = 1~~(1 \leq j, j' \leq J; j
\ne j')\\
~~~~~~~~~~~~~~~~~~~~~~~~~~~~~~~~~~~~~~~~~( \mbox{otherwise} ),
\end{cases}
\label{CostFunction}
\end{eqnarray}
where $\mathbf{A}'$ is constructed from a set $\mathcal{W} + \{
\mathit{path} \}$,
and this cost function is used in getCostMin($\mathcal{P}$).
Once the mutual coherence of the constructed matrix becomes less than 1.0,
this algorithm terminates.
STEP 2 cannot directly select a path depending on the number of nodes
over the path.
Therefore, the proposed routing matrix construction algorithm
pays attention to the interval factor rather than the traffic factor.


\begin{algorithm}
\caption{Proposed Routing Matrix Construction Algorithm}
\label{algorithm}
\begin{algorithmic}
\REQUIRE Network Topology and $s$.
\ENSURE Routing Matrix $\mathbf{A}$.
\STEP {\bf 1} : \underline{Search for path candidates}
\FORALL{$v \in \mathcal{V} \setminus s$}
\STATE $\mathcal{P}_{{\rm disjoint}} := {\rm NodeDisjointAlgorithm}(s,v)$.
\FORALL{$\mathit{path}^{(a)} \in \mathcal{P}_{{\rm disjoint}} (a = 1,2
\cdots | \mathcal{P}_{{\rm disjoint}} |) $}
\FORALL{$\mathit{path}^{(b)} \in \overline{ \mathcal{P} }_{{\rm
disjoint}} (b = 1,2 \cdots | \overline{\mathcal{P}}_{{\rm disjoint}}|)$}
\STATE $\mathcal{P}_{{\rm all}} := \mathcal{P}_{{\rm all}} \cup \{
\mathit{path}^{(a)} + \mathit{path}^{(b)} \}$.
\ENDFOR
\ENDFOR
\ENDFOR
\STEP {\bf 2} : \underline{Selection of measurement paths}
\WHILE{ $F_{\mu} (\mathbf{A}) \geq 1.0$ }
\STATE $\mathit{path}^{({\rm min})} := {\rm getCostMin} \left(
\mathcal{P}_{{\rm all}} \setminus \mathcal{W} \right) $.
\STATE $\mathcal{W} := \mathcal{W} \cup \{ \mathit{path}^{({\rm min})} \}$.
\STATE Construct $\mathbf{A}$ from set $\mathcal{W}$.
\ENDWHILE
\RETURN $\mathbf{A}$.
\end{algorithmic}
\end{algorithm}


\begin{figure*}[!t]
\centerline{
\subfloat[Network topology]{\includegraphics[clip,
width=0.6\columnwidth]{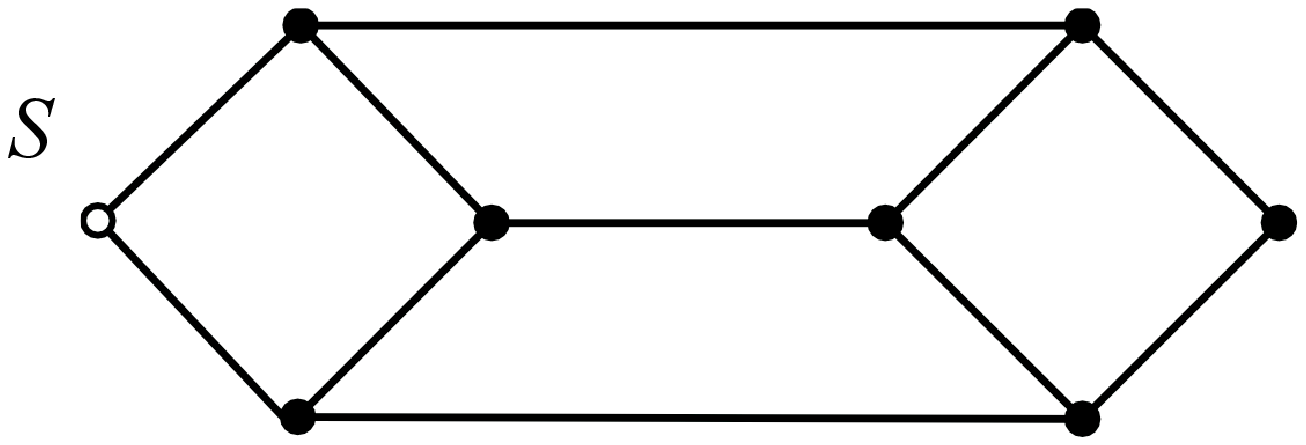}
\label{fig:label-A}}
\hfil
\subfloat[Measurement
path~($\mathit{path}_{s}^{(1)}$)]{\includegraphics[clip,
width=0.6\columnwidth]{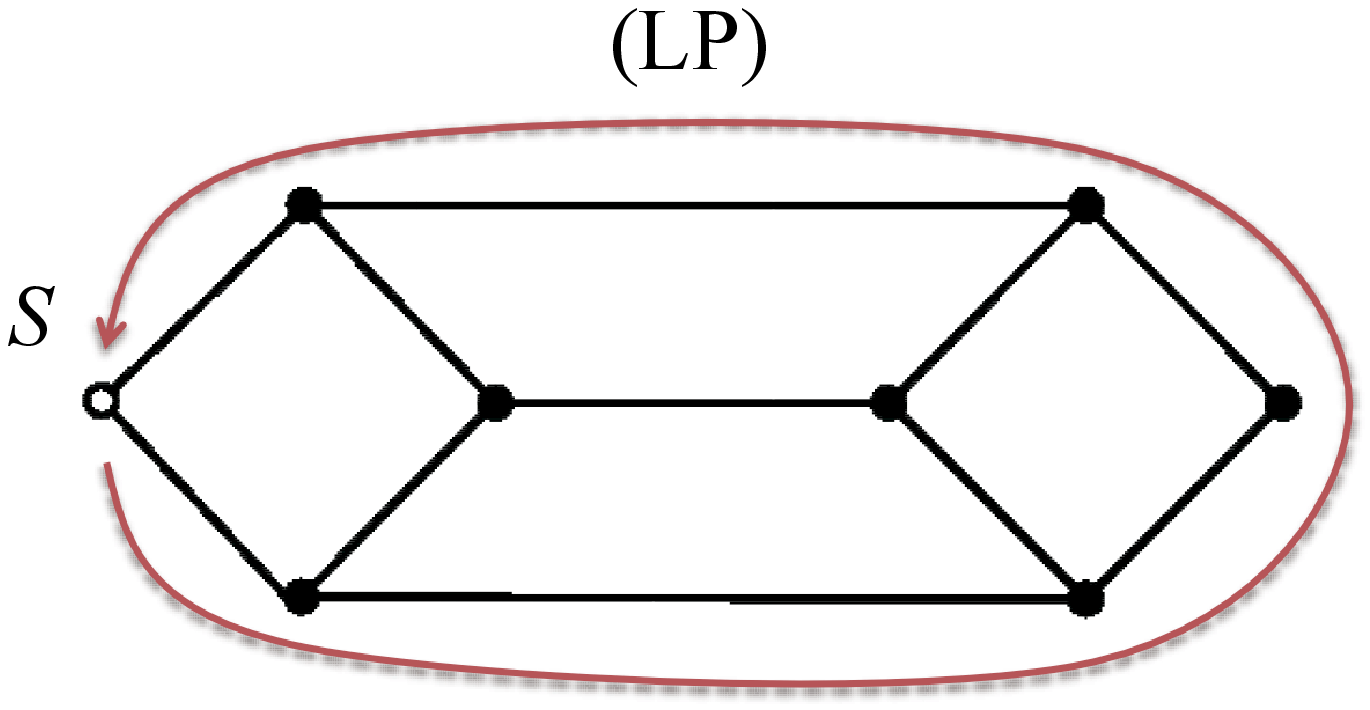}
\label{fig:label-B}}
\hfil
\subfloat[Measurement
path~($\mathit{path}_{s}^{(2)}$)]{\includegraphics[clip,
width=0.6\columnwidth]{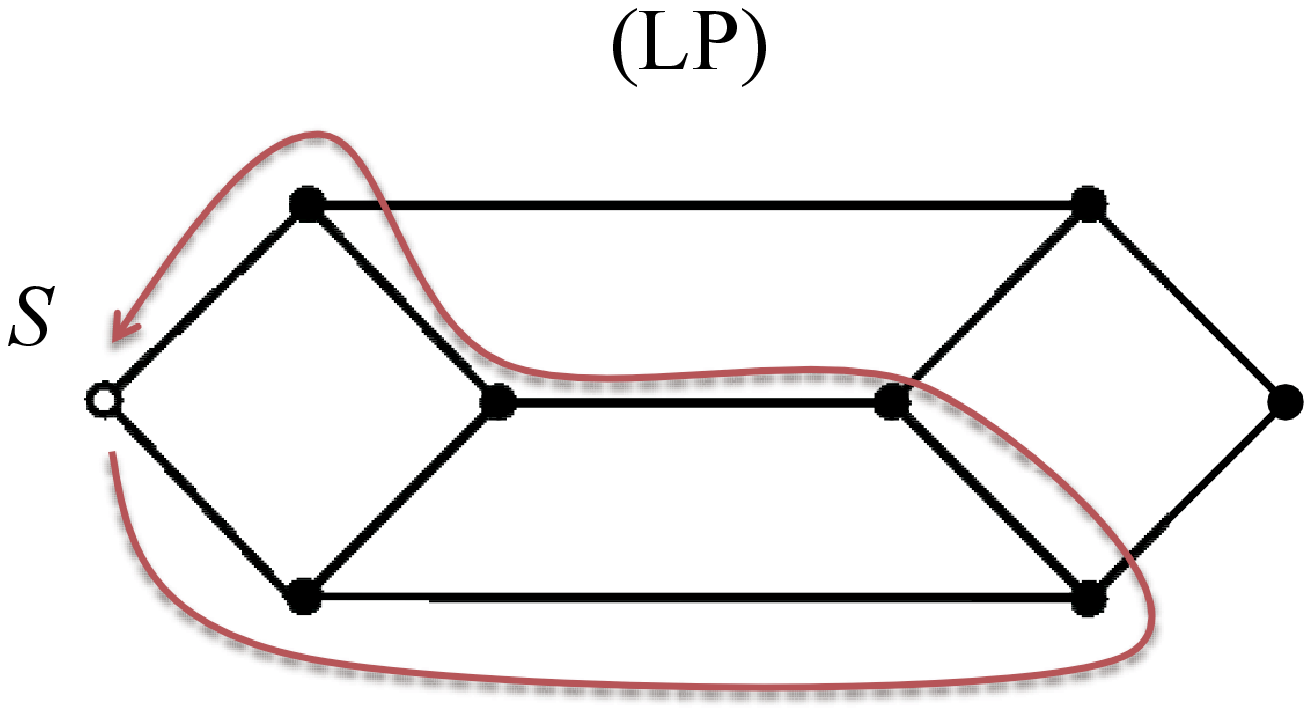}
\label{fig:label-C}}
}

\centering

\centerline{
\subfloat[Measurement
path~($\mathit{path}_{s}^{(3)}$)]{\includegraphics[clip,
width=0.6\columnwidth]{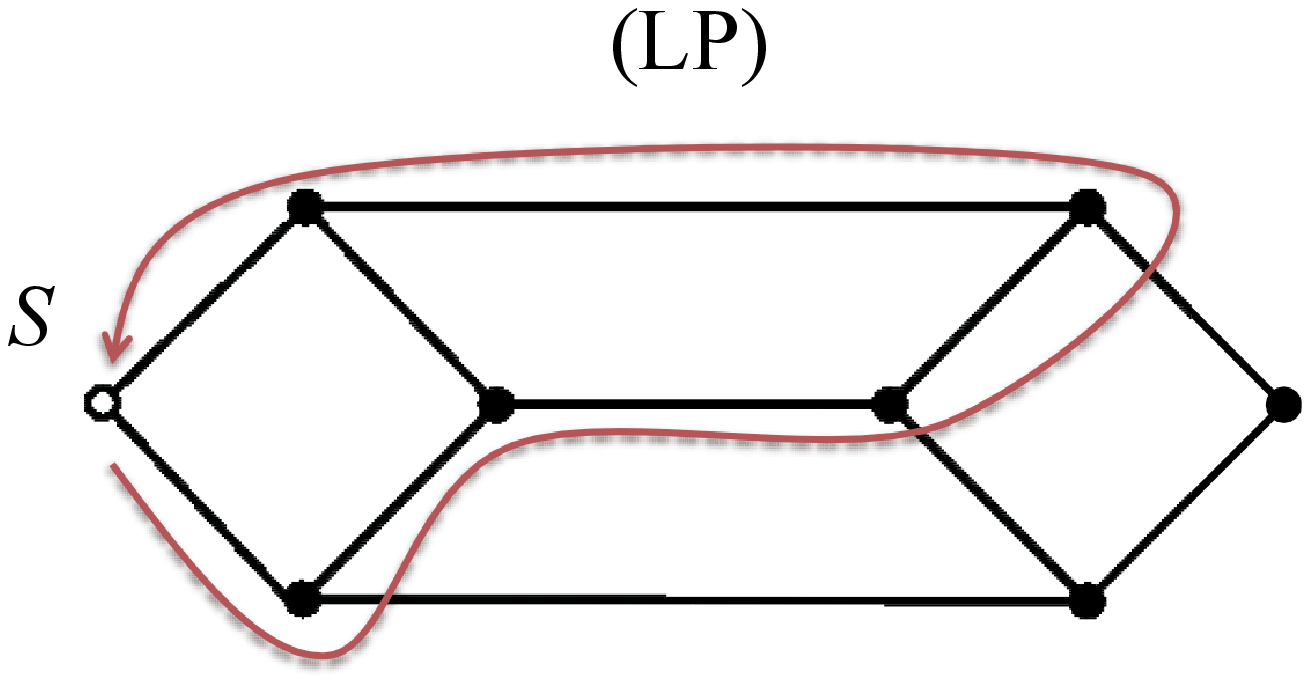}
\label{fig:label-A}}
\hfil
\subfloat[Measurement
path~($\mathit{path}_{s}^{(4)}$)]{\includegraphics[clip,
width=0.6\columnwidth]{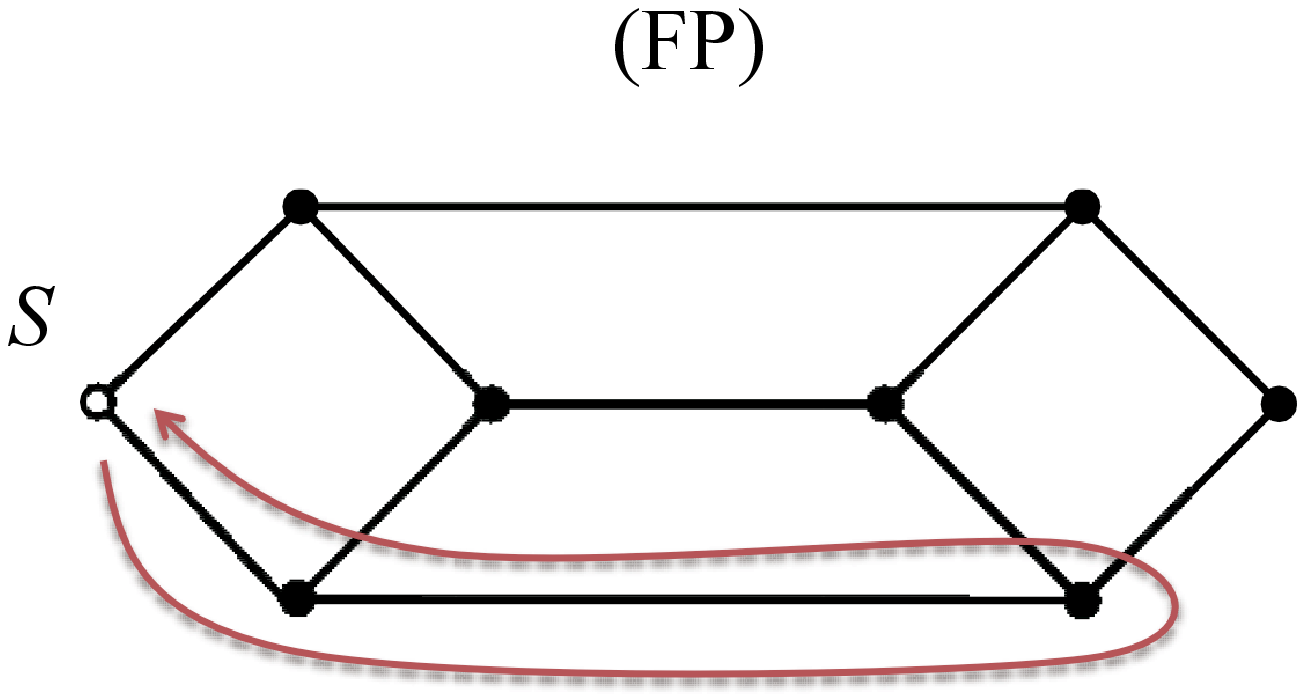}
\label{fig:label-B}}
\hfil
\subfloat[Measurement
path~($\mathit{path}_{s}^{(5)}$)]{\includegraphics[clip,
width=0.6\columnwidth]{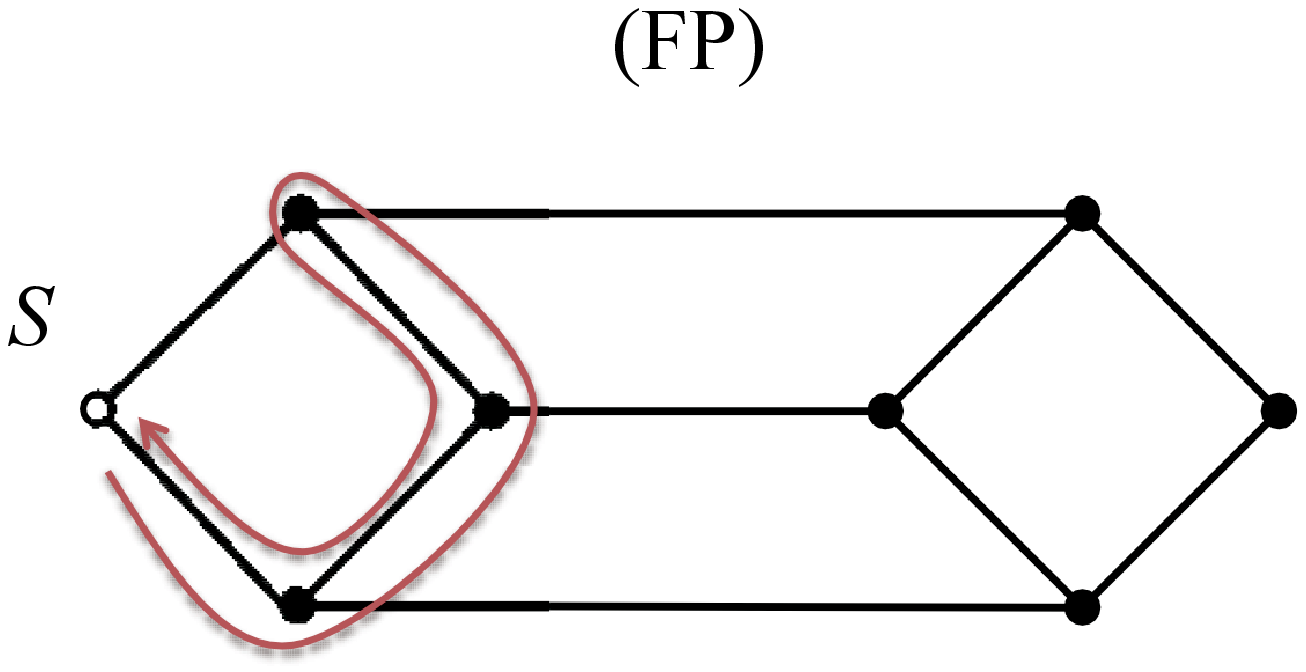}
\label{fig:label-C}}
}
\caption{Network topology (8 nodes and 16 links) and the measurement
paths constructed by the proposed algorithm.}
\label{fig:topology}
\end{figure*}

\section{Performance Evaluation}
\label{sec:evaluation}
In this section, we discuss the following items:
\begin{itemize}
\item
Can the proposed algorithm (Algorithm~\ref{algorithm}) construct a fully
adequate routing matrix?

\item
Can
a routing matrix constructed by the proposed algorithm actually identify
a bottleneck
link in a network where only one bottleneck link exists?

\item
How does a routing matrix with smaller interval and traffic factors
behave in a network
where several bottleneck links exist?
\end{itemize}

Fig.~\ref{fig:topology}(a) shows the network topology with $8$ nodes and
$11$ links used
for performance evaluation by computer simulation,
where there is only one transceiver node $s$.
We assume that
the delay of a bottleneck link is constant
with $x^{(B)}$
whereas that of a normal link
denoted by $x^{(N)}$ is
independent and identically distributed (i.i.d.)
with average $\alpha_{x^{(N)}}$
and standard deviation $\sigma_{x^{(N)}}$.
In this paper,
we also assume that all the nodes are wirelessly connected
so $x^{(N)}$ is Gaussian-distributed~\cite{Noh2007} with
$\alpha_{x^{(N)}} =15$ msec and
$\sigma_{x^{(N)}} =3$ msec~\cite{Zeng2009,Liu2013}.

\begin{table}[t]
\begin{center}
\caption{Routing Matrices}
\begin{tabular}{|c|c|c|c|c|}\hline
\raisebox{0.9ex}{Matrix} &
\raisebox{0.9ex}{Size} &
\shortstack{Entrywise\\Matrix Norm} &
\shortstack{LP:FP\\Numbers} &
\shortstack{Number of \\ Candidates}\\ \hline
$\mathbf{P}_{1}$ & $5 \times 11$& $30$ & $3:2$ &$18$paths \\ \hline
$\mathbf{P}_{2}$ & $5 \times 11$& $28$ & $4:1$ &$18$paths \\ \hline
$\mathbf{P}_{3}$ & $4 \times 11$& $28$ & $3:1$ &$78$paths \\ \hline
\end{tabular}
\label{matricesP}
\end{center}
\end{table}

First, Table~\ref{matricesP} shows the constructed three routing
matrices whose
mutual coherences are less than $1$.
In Table~\ref{matricesP},
$\mathbf{P}_{1}$ is constructed by the proposed algorithm composed
of STEP 1 and STEP 2 (the measurement paths are shown in
Figs.~\ref{fig:topology}(b)-(f)),
$\mathbf{P}_{2}$ is constructed by a greedy search from the path
candidates listed by STEP 1
(instead of STEP 2, the paths are selected from all combinations of the
path candidates by STEP 1,
which minimizes the interval and traffic factors), and
$\mathbf{P}_{3}$ is also constructed by a greedy search from path
candidates listed by STEP 1
and additional FP candidates
(all FPs are added to the path candidates by STEP 1 and then the paths
are selected
from all combinations of the increased path candidates, which minimize
the interval and traffic factors).
It is impossible for the proposed algorithm
to always select the paths which really minimize the interval and
traffic factors
due to its one-by-one policy (in STEP~$2$),
on the other hand,
the greedy search-based algorithms can always select the optimum set of
paths from all combinations of paths.
Comparing $\mathbf{P}_{1}$ and $\mathbf{P}_{2}$,
the proposed algorithm composed STEP 1 and STEP 2 constructs the routing
matrix whose
traffic factor is a little larger,
and comparing $\mathbf{P}_{2}$ and $\mathbf{P}_{3}$,
the number of path candidates by STEP 1 seems insufficient.
However,
when the size of network is large,
for the case where $I$ measurement paths are selected from $N$ candidates,
the proposed algorithm calculates the cost function ($IN+I(I-1)/2$) times,
whereas the greedy search-based algorithms lead to {\em combinatorial
explosion}.
So, taking into consideration that
computational complexity of the proposed algorithm is much lower than
that of the greedy search-based algorithm,
it can be concluded that the proposed algorithm can efficiently
construct a fully adequate routing matrix.

\begin{figure}[ttt]
\centering
\includegraphics[width = .7\linewidth]{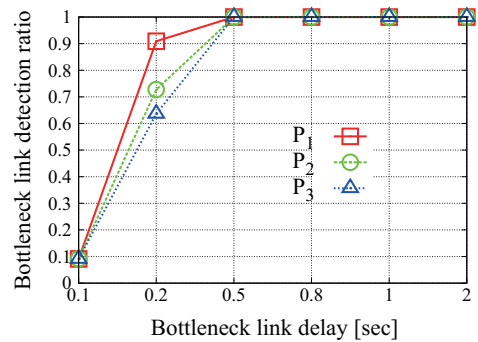}
\caption{Bottleneck link detection ratio vs. bottleneck link delay
$x^{(B)}$
(Number of bottleneck links $k = 1$).}
\label{resultBLDR}
\end{figure}

The termination of the proposed algorithm is guaranteed since
as the number of measurement paths increases,
the mutual coherence of the constructed routing matrix monotonously
decreases.
While mutual coherence can provide a guarantee of the recovery of
exactly sparse vectors,
the link delay vector $\mathbf{x}$ is approximately sparse in the model
for performance evaluation.
Therefore, to confirm whether or not the bottleneck link detectability
of the reflective network tomography scheme is consistent with the
meaning of the
mutual coherence.
So, we assumed that there is a bottleneck link in the network, that is,
we set the number of bottleneck links $k$ to $1$ in the computer
simulation.
Here,
we also define {\em bottleneck link detection ratio}
which is defined as the number of correctly detected
bottleneck links divided by the total number of given bottleneck
links.
Fig.~\ref{resultBLDR} shows the bottleneck link detection
ratio versus the bottleneck link delay for $k = 1$.
Although the link delay vector $\mathbf{x}$ is not exactly sparse,
as the bottleneck link delay $x^{(B)}$ becomes larger,
the bottleneck link detection
ratios of the three routing matrices approaches $1.0$.
This means that,
if the bottleneck link delay $x^{(B)}$ is fully larger,
the link delay vector $\mathbf{x}$ can be regarded
approximately as a sparse vector,
and the mutual coherence can also guarantee the recovery of
approximately $1$-sparse vectors.
Thus, it can be concluded that the proposed scheme can effectively
detect a bottleneck link.

\begin{figure}[ttt]
\centering
\includegraphics[width = .7\linewidth]{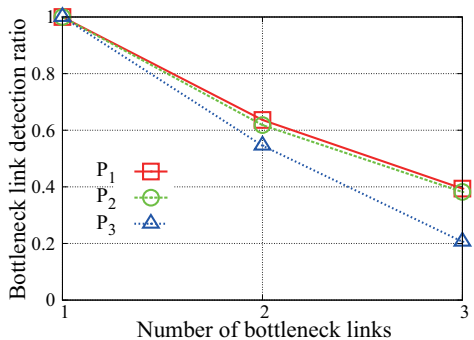}
\caption{Bottleneck link detection ratio vs. number of
bottleneck links $k$
(Bottleneck link delay $x^{(B)}=1.0$).}
\label{result2}
\end{figure}

Finally,
we evaluated the matrices $\mathbf{P}_{1},\mathbf{P}_{2},\mathbf{P}_{3}$
in the network with the number of the bottleneck links $k = 1,2,3$.
Fig.~\ref{result2} shows the bottleneck link detection
ratio versus the number of
bottleneck links,
where we set $x^{(B)}$ to $1$ sec,
which corresponds to
about $66.6$ times as large as $\alpha_{x^{(N)}}$.
For $k \ge 2$, all the bottleneck link detection
ratios fall down sharply.
This is because that the algorithms
introduced here all try to construct routing matrices with smaller interval
and traffic factors, which have worse impact on the bottleneck link
detectability
of $k \ge 2$
(the proposed algorithm composed of STEP 1 and STEP 2 pays attention only
to reducing the interval factor, but it also results in reduction
of the traffic factor).
Therefore,
for a network with the possibility that several bottleneck links arise
simultaneously,
we need to redesign the termination condition and cost function.

\section{Conclusion}
\label{conclusion}

In this paper, according to the types of end-to-end path measurements
acquisition,
we classified network tomography into transmissive and reflective schemes
and proposed a new reflective network tomography
with their advantageous characteristics over conventional transmissive
network tomography.
We proposed a simple reflective routing matrix construction algorithms
composed of two steps,
and by computer simulation we showed that it can effectively
construct an adequate routing matrix guaranteeing a designed
bottleneck link detectability of $k=1$.

Some technical
issues remain in the proposed scheme.
First, we have to propose a better routing matrix construction algorithm,
and evaluate reflective network tomography on larger networks.
We also have to
propose an adaptive network tomography
to take advantage of reflective characteristic.
Since these issues are beyond
the scope of this paper, we leave them as future works.

\section*{Acknowledgment}
This work was supported in part by the Japanese Ministry
of Internal Affairs and Communications in R\&D on Cooperative
Technologies and Frequency Sharing Between Unmanned
Aircraft Systems (UAS) Based Wireless Relay Systems and
Terrestrial Networks.



%

\end{document}